\documentclass[prb,twocolumn,amsmath,amssymb,showpacs]{revtex4}
\usepackage{graphicx}
 
\begin{document}

%
%

\title{Large-Scale Monte Carlo Study of a Realistic Lattice Model for $\rm Ga_{\it 1-x}Mn_{\it x}As$.}

\author{Yucel Yildirim$^{1,2}$, Gonzalo Alvarez$^3$, Adriana Moreo$^{1}$, 
and Elbio Dagotto$^{1}$}
 
\address{$^1$Department of Physics and Astronomy,University of Tennessee,
Knoxville, TN 37966-1200 and
\\Oak Ridge National Laboratory,Oak Ridge,
TN 37831-6032,USA, }
 
\address{$^2$Department of Physics, Florida State University, Tallahassee, FL 32306, USA,}

\affiliation{$^3$ Computer Science and Mathematics Division, Oak Ridge
National Laboratory, Oak Ridge, TN 37831-6032, USA.}

\date{\today}

\begin{abstract}

The properties of Mn-doped GaAs are studied at several doping levels 
and hole compensations, using a real-space Hamiltonian on an fcc lattice 
that reproduces the valence bands of undoped GaAs. Large-scale
Monte Carlo (MC) simulations on a Cray XT3 supercomputer,
using up to a thousand nodes, were needed to make this effort possible.
Our analysis considers both 
the spin-orbit interaction and the random distribution of the Mn 
ions. The hopping amplitudes are functions of the GaAs Luttinger parameters. 
At the coupling $J$$\sim$1.2~eV deduced from photoemission experiments, 
the MC Curie temperature and the shape of the magnetization curves are 
in agreement with experimental results for annealed samples. 
Although there are
sizable differences with mean-field predictions, 
the system is found to be closer to a hole-fluid regime than to 
localized carriers.

\end{abstract}
 
 
\maketitle


{\it Introduction}. The study of diluted magnetic 
semiconductors (DMS) is a 
rapidly expanding area of research\cite{OHN96,dietl,RMP,MacDonald,Thomas}  
 triggered by the discovery of high Curie 
temperatures ($T_C$) in some of these materials, and by their
potential role in spintronics devices.\cite{zutic}
Experimentally, much progress in the study of DMS was recently 
achieved after annealing techniques 
reduced the rates of compensation, allowing for 
higher $T_C$'s and for the intrinsic properties of DMS materials to be 
carefully investigated.\cite{Potash} 
Theoretical studies have focused on two idealized scenarios:
(1) Impurity-Band (IB) models;\cite{IB}
and (2) Hole-Fluid (HF) models (based on
mean-field approximations),\cite{dietl,macdonald}
which produce dramatic differences in most physical
quantities. 

These rather different views (1) and (2) of the same problem show that
the use of unbiased computational techniques, such as lattice
MC simulations, is crucial for progress in the modeling of DMS materials since
it is unlikely that the coupling constants and densities locate the DMS 
systems exactly in one extreme or the other. Flexible methods that can 
interpolate between
both limits within the same formalism are needed.
However, while
considerable progress has been achieved in numerical
studies of simplified models\cite{nosotros}, the
complexity of the real problem (involving several bands on an fcc lattice) has
prevented its detailed analysis until now.

Here, we report a comprehensive  
numerical Monte Carlo study of a realistic lattice model for Mn-doped GaAs, including spin-orbit coupling, as
well as the effects of random Mn doping. 
This large-scale computational effort was possible by using the Cray XT3 supercomputer operated 
by the National Center for Computational Sciences. Our simulations made use of 
up to 1,000 XT3 nodes. Parallelization was used in different ways: 
(i) to study different regions of parameter space (densities, couplings), and 
(ii) to average over different 
configurations of Mn locations. 
In all cases, the use  of hundreds of processors in a single parallel run 
poses several technical challenges that are best handled 
by supercomputers with low latency and scalability, rather than by conventional clusters of PC's. 
In fact, this study would have taken several 
years without access to 
a supercomputer with thousands of processors, such as the Cray XT3.

{\it The Model}. 
We have constructed a real-space fcc-lattice Hamiltonian whose kinetic-energy term 
maps into the Luttinger-Kohn model,\cite{KL} when $k$-space Fourier transformed and at $k\rightarrow 0$. 
As a consequence, the hopping amplitudes are functions of (tabulated) Luttinger parameters, and thus they
are precisely known.\cite{Lut} To incorporate the spin-orbit (SO) interaction, we work in
the $|j,m_j\rangle$ basis, where $j$ can be $3/2$ or $1/2$ (since we consider 
the $p$ orbitals, $l$=1, relevant at the $\Gamma$ point of the GaAs valence 
band). Consequently, there 
are 6 possible values for $m_j$, indicating that this is a fully 6-orbital 
approach, arising from the 3 original $p$ orbitals and the 2 hole spin 
projections. The Hamiltonian is formally given by
$$
{\rm H={1\over{2}}
\sum_{{\bf i,\mu,\nu},\alpha,\alpha',a,b}(t_{\alpha a,\alpha' b}^{\bf\mu\nu}
c^{\dagger}_{{\bf i},\alpha a}c_{{\bf i+\mu+\nu},\alpha' b}+h.c.)}
$$
$$
+\Delta_{SO}\sum_{{\bf i},\alpha}c^{\dagger}_{{\bf i},\alpha{\tilde1\over{2}}}
c_{{\bf i},\alpha{\tilde1\over{2}}}+
{J\sum_{{\bf I}}{\bf s_I}\cdot{\bf S_I}},
\eqno(1)
$$
\noindent where ${\rm a,b}$ take the values ${1\over{2}}$, ${3\over{2}}$
(for $j$=$3/2$), or ${\tilde1\over{2}}$ (for $j$=$1/2$), and 
$\alpha$ and $\alpha'$ can be $1$ or $-1$. The Hund term describes the 
interaction between the hole spins ${\bf s_I}$ (expressed in 
the $|j,m_j\rangle$ basis\cite{6x6}) and the spin of the localized Mn ion 
${\bf S_I}$.
The latter is considered classical ($|{\bf S_I}|$=1),\cite{foot} since it is 
large $S=5/2$.\cite{spin} 
{$\bf\mu+\nu$} are the 12 
vectors indicating the 12 nearest-neighbor (NN) sites of each ion located 
at site {\bf i}, while
{\bf I} are random sites in the fcc lattice. 
$\Delta_{SO}$=$0.341$ eV is the spin-orbit 
interaction strength\cite{cardona} in GaAs. The hopping parameters, 
$t_{\alpha a,\alpha' b}^{\bf\mu\nu}$, are complex numbers, whose real and 
imaginary parts are functions of the Luttinger parameters.\cite{6orbs}
\begin{figure}[thbp]
\begin{center}
\includegraphics[width=8cm,clip,angle=0]{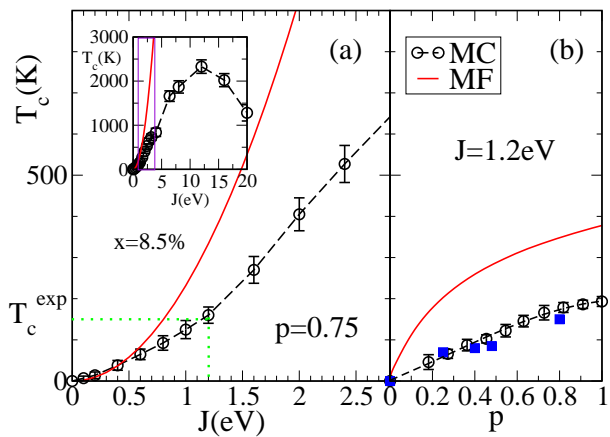}
\vskip 0.3cm
\caption{(color online)(a) Curie temperature vs. $J$, for $x$=$8.5\%$ and $p$$\approx$$0.75$. The MC 
results are indicated by circles, while the continuous line is the MF prediction.\cite{dietl,priv}
$Inset$: MC results for larger values of $J$ to observe the crossover toward 
a localized picture. Vertical lines indicate the experimentally acceptable 
range of $J$. (b) MC calculated $T_C$ vs. $p$, at $x$=8.5\%.
The blue dots are experimental results,\cite{po2} 
and the solid line the MF prediction.}
\label{TcvsJ}
\end{center}
\end{figure}

Equation (1) can be studied with standard MC techniques 
for systems involving fermions and classical spins. These conceptually simple
methods
were already described in much detail in previous DMS 
investigations\cite{nosotros}
and also in manganese oxide studies,
thus details will not be repeated here.
Numerically, we analyze clusters that contain $N=N_xN_yN_z$ unit cells ($N_i$ 
is the number of cells along the spatial direction $i$) of 
side $a_0$ ($a_0$=$5.64{\rm \AA}$,\cite{cardona} is the $GaAs$ cubic 
lattice parameter).
Since in an fcc lattice there are 4 ions associated to each cell,
the total number of $Ga$ sites is
$N_{Ga}$=$4N$. Since there are 6 single fermionic states per site, 
the diagonalization of a $6N_{Ga}\times 6N_{Ga}$ matrix is needed at each step of the MC 
simulation, which demands considerable computational resources for large 
enough clusters. The 
diagonalization can be performed exactly for values of $N_{Ga}$ up to 500. 
We show below that lattices with $N_i=4$ (256 sites) are 
large enough to study Mn dopings $x$ and compensations $p$
in the range of interest, with sufficient
precision for our purposes. 
Nominally there should be one 
hole per Mn ion, but $p$ can be smaller than 1 due to hole trapping defects, 
thus $p$ and $x$ are considered independent in this study.

{\it Results}. The highest $T_C$ experimentally observed in 
bulk Ga$_{1-x}$Mn$_x$As is $\sim 150~K$, at $x$=$8.5\%$ 
and $p$$\approx$$0.7$\cite{po2,tc}. 
The system is 
metallic, and the magnetization vs. temperature displays mean-field
behavior.\cite{po2} In Fig.~1a, we present the (MC calculated) $T_C$ as a 
function 
of the coupling $J$, at 
$x$=$8.5\%$ and $p$$\approx$$0.75$. 
The results shown were obtained on lattices containing 256 
(Ga,Mn) ions, and using an average over at least 5 different disorder 
configurations (only small differences 
were observed among the Mn configurations). Results
for lattices with up to 500 sites have also been calculated for some 
parameters (see below).
At the realistic $J$=$1.2~eV$\cite{oka}, Fig.~1a shows that the critical
temperature is $T_C$=$155\pm 20 K$. Since
$J$ is not accurately known, this excellent agreement with experiment\cite{po2}
could be partially fortuitous, but at least the results indicate that a reasonable
quantitative estimation of the real $T_C$ can be made via MC simulations of
lattice models.
The solid line in the figure corresponds to the
MF results.\cite{dietl,dietlprb,priv} The quantitative
MF-MC agreement
at small $J$ provides a strong test of the reliability of the present MC
approach.
At $J$=1.2~eV, the MF $T_C$ is $\sim$300~K, showing that at these
couplings and densities appreciable
differences between MC and MF exist: the fluctuations considered in the MC
approach cannot be neglected.
The inset of Fig.~1 demonstrates that eventually for very
large values of $J$ the MF approximation breaks down, as expected.
The MC simulations show that
$T_C$ reaches a maximum for $J\approx 12eV$, of the order of the
carriers bandwidth,
and then it decreases due to the tendency of  holes
toward strong localization. This ``up and down'' behavior can only be obtained with lattice MC
simulations valid at arbitrary values of $J$.\cite{nosotros}

At $J$$\sim$$1.2eV$ the system is
closer to a hole-fluid than a localized regime as suggested
by the magnetization vs. $T$ curve, displayed
in Fig.~2a. This curve has Curie-Weiss shape in qualitative agreement with both
experimental results \cite{po2} and previous MF
calculations.\cite{dietlprb,priv,macdonald} This qualitatively correct shape of the magnetization
curve was not obtained in previous lattice MC simulations.\cite{nosotros}
Size effects are mild as it can be seen in
Fig.~3a where data for
magnetization vs. $T$ are presented for $x=8.5\%$,
$p\approx 0.75$, and
$J=1.2 eV$ in lattices with $(N_x,N_y,N_z)=(4,3,3)$, $(4,4,4)$, $(5,4,4)$, and
$(6,4,4)$, i.e., with $N_{Ga}$=144, 256, 320, and 384. Results for N=500 were
obtained for lower doping (see Fig.~2b). Considering
together the results for the different size clusters the estimated $T_C\approx 155 \pm 15$
K is still in agreement with  experiments (and also with the 256 sites results). 
Regarding the Curie-Weiss (CW) shape of the
magnetization curve, we
have phenomenologically observed that the finite spin-orbit coupling plays a
crucial role in this
respect. In Fig.~3b we show the magnetization vs. $T$ for $x=8.5\%$,
$p=0.75$, and $J=1.2 eV$ for $\Delta_{SO}=0.34eV$ (squares) and
$\Delta_{SO}=0$ (circles): only the nonzero SO coupling produces CW behavior.
 In addition, we have noticed that the
CW shape is also missing with the 4 orbital (with $j$=3/2) model that results
in the limit $\Delta_{SO}\to\infty$.\cite{6orbs} This indicates the important
role that a realistic representation of the valence band plays in properly
describing the thermodynamic observables.
  
\begin{figure}[thbp]
\begin{center}
\includegraphics[width=8cm,clip,angle=0]{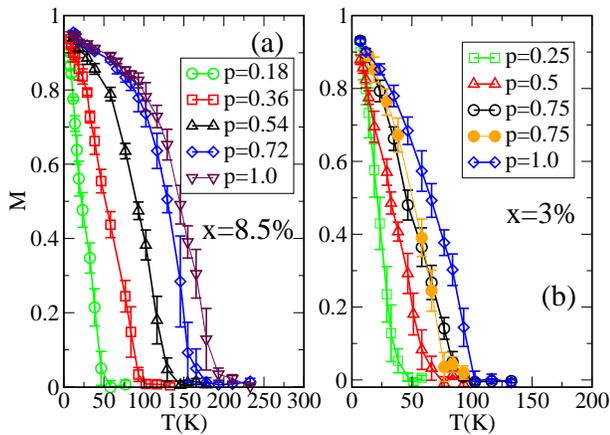}
\vskip 0.3cm
\caption{(color online)(a) Magnetization M vs. $T$, for $x$=$8.5\%$ and several $p$'s 
(indicated), using a 256 sites lattice (open symbols). 
Averages over 5 Mn-disorder configurations are shown.  
(b) Same as (a), but for $x$=3\%. Close circles are
results for a 500 sites lattice.
 The magnetization is measured as 
$\cal M=\sqrt{{\bf M}\cdot{\bf M}}$,
with ${\bf M}$ the vectorial magnetization. As a consequence, 
for fully disordered spins, $\cal M$ is
still nonzero due to the ${\bf S_I}^2$=1 contributions, causing a finite 
value at large temperatures (${\cal M}(T\to\infty)=1/\sqrt{xN_{Ga}}$)
unrelated to ferromagnetism. Thus, we plotted
${\rm M}=({\cal M}-{\cal M}(T\to\infty))/(1-{\cal M}(T\to\infty))$, i.e. the
background was substracted.}
\label{Fig2}
\end{center}
\end{figure}
The charge distribution in the cluster provides interesting information.
In the HF scenario, the charge is assumed to be uniformly distributed
while in the IB picture the charge is strongly
localized. Fig.~3c indicates that in the realistic regime
with $J=1.2$ eV, $x=8.5\%$, and $p\approx 0.75$ the charge is fairly uniformly
distributed. The slightly darker points correspond
to the sites where the Mn are located. They have charge of the
order of 20\% above the MF value defined as $n_{MF}$=$n/N_{Ga}$
(with $n$ the number of holes).
As shown in Fig.~3d, charge localization occurs when $J$ is increased to
large values such as 12 eV. The dark circles at the Mn sites have
charge intensities about 20 times the MF value,  with very little
charge found away from the impurities.
\begin{figure}[thbp]
\begin{center}
\includegraphics[width=8cm,clip,angle=0]{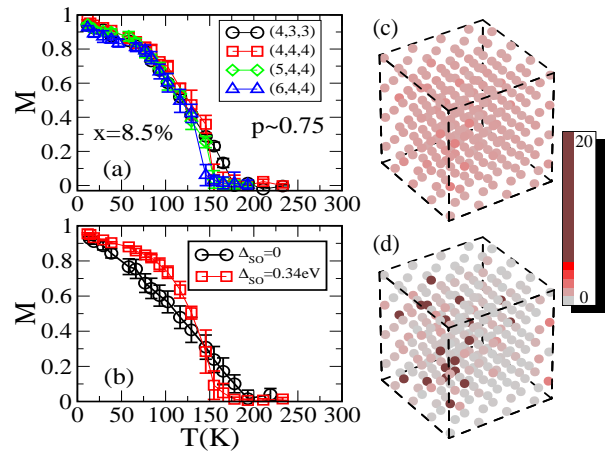}
\vskip 0.3cm
\caption{(color online)(a) M (defined as in Fig.~2) vs $T$ for different
lattice sizes for $x=8.5\%$, $p\approx 0.75$, and $J=1.2 eV$; 
(b) M vs. $T$ for the same parameters
as in (a) on a 256 sites lattice with (without) spin-orbit interaction 
indicated by the squares (circles); 
(c) Charge density normalized to the MF value (see text),
for $x=8.5\%$, $p$$\approx$
$0.75$, $T$=10K, on a 256-sites cluster for $J$=$1.2eV$.
The color intensity is proportional to the charge density (see 
scale). (d) Same as (c), but for
$J$=$12 eV$.}
\label{Fig3}
\end{center}
\end{figure}

The lack of charge localization effects in the ordered state is 
concomitant with
the absence of a notorious impurity 
band in the density-of-states (DOS) (Fig.~4a). 
Increasing $J$, an IB regime is eventually observed, given confidence that the
study is truly unbiased. This occurs for $J\approx 4eV$ and beyond, 
with the IB becoming totally
detached  from the valence band at $J\approx 16eV$. Figures~3c, d and 
4a show that 
the degree of spatial hole localization is correlated with the 
development of the IB.
In addition, when the  holes are localized,
the M vs. $T$ curves
present substantial deviations from MF behavior (not shown), with a different concavity
as that of Fig.~2a.\cite{nosotros}
\begin{figure}[thbp]
\begin{center}
\includegraphics[width=8cm,clip,angle=0]{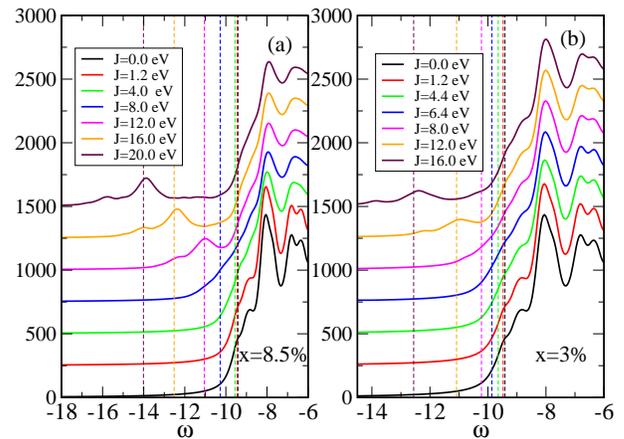}
\vskip 0.3cm
\caption{(color online)(a) Density-of-states, for $x$=$8.5\%$, $p$$\approx$$0.75$, and several
$J$'s. The dashed vertical lines indicate the position of 
the chemical potentials. (b) Same as (a) but for $x=3\%$.}
\label{Fig4}
\end{center}
\end{figure}

It is well known that the carrier density in DMS is strongly dependent on 
sample preparation. Due to defects, $p$ is much smaller 
than 1 in most samples. 
In Fig.~1b, the MC calculated
$T_C$ vs $p$ for $x$=$8.5\%$ and $J$=$1.2eV$ is shown. 
$T_C$ increases
with $p$, and it reaches a maximum at $p$$\sim$$1$ as in previous 
theoretical,\cite{dietl,nosotros}  
and experimental results\cite{Potash,po2,exp,exp1}  also shown in the figure.
The agreement between MC results and experiments is once again quite satisfactory. The
figure suggests that if $p$=1 were reachable experimentally, 
then $T_C$ could be
as high as $\sim$200~K.
We also observed qualitative 
changes in the magnetization curve varying $p$ (Fig.~2a): as $p$ is 
reduced, the magnetization changes from a Brillouin form to an approximately linear shape with $T$. 
In fact, the observed 
$p$ dependence is once again similar as found experimentally, with $p$ modified
by annealing;\cite{po2} the Mn disorder 
plays a more dominant role when the 
number of holes is reduced. In spite of the deviations from
MF behavior at small $p$,  we could not observe a clear IB
as $p$ was reduced, for the coupling $J$ used in our analysis. 

Consider now the low
Mn-doping regime.
The latest experimental results suggest that (Mn,Ga)As should
still be metallic at $x$=3\%.\cite{exp1} A metal-insulator transition is
expected
at $x\le 1\%$,\cite{mit} but MC studies for such very small dopings need
much larger clusters than currently possible.
Our MC simulations indicate that at low Mn-doping,
the dependence of $T_C$ with $J$ is
similar as in the higher doping case.

No clearly formed IB is observed in the $x$=3\% DOS
displayed in Fig.~4b: the IB's are formed for similar values of $J$ with increasing $J$,
at both $x$=3\% and 8.5\%, as deduced from an analysis of fermionic eigenvalues for
spin-ordered configurations.\cite{6orbs,coul} 

{\it Summary.}
Here, a MC study of an fcc lattice model for DMS compounds,
including the realistic valence bands of GaAs, the spin-orbit interaction,
and the random distribution of Mn dopants has been presented.
The use of the Cray XT3 at ORNL
made this effort possible. The results show magnetizations and
$T_C$'s in reasonable
agreement with experiments.
The simulations show that the carriers tend to
spread over the entire lattice, and they reside in the valence band
at realistic couplings, qualitatively in agreement with MF
\cite{dietl,macdonald} and first principles \cite{Thomas}
calculations in the same parameter regime, as well as with experimental data on
annealed samples.\cite{exp1,Potash} However, an IB band populated by a fraction $1-p$
of trapped holes that do not participate in the transport properties is not
ruled out by our results.
The MC method described here opens a new semi-quantitative window for theoretical
research on the properties
of DMS materials.

{\it Acknowledgments.} 
We acknowledge discussions with T. Dietl, A. MacDonald,
J. Sinova, T. Schulthess, F. Popescu and J. Moreno. Y.Y., A.M., and E.D.
are supported by NSF under grants DMR-0443144 and DMR-0454504.
G.A. is sponsored by the Division of Materials
Sciences and Engineering, Office of Basic Energy Sciences, U.S. Department
of Energy, under Engineering,
Office of Basic Energy Sciences, DOE, under
contract DE-AC05-00OR22725 with ORNL, managed and operated by UT-Battelle, LLC.


\begin{thebibliography}{}
\bibitem{6x6} These are $6 \times 6$ matrices; see 
Eq.(A.10) (in Ref.\onlinecite{macdonald}) and \onlinecite{6orbs}.
\bibitem{finite} N=500 results were obtained for lower doping 
(Fig.~2b).
\end{thebibliography}
\end{document}